\documentclass[a4paper,11pt,twoside]{article}
\usepackage{epsfig,graphics,float}
\usepackage{subfig}
\usepackage{geometry}
\usepackage{amsmath}
\usepackage{amssymb}
\usepackage{color}
\usepackage{graphicx}
\usepackage{floatflt}
\usepackage{rotating}
\usepackage{indentfirst}
\usepackage{amsmath}
\usepackage{amssymb}
\usepackage{mathrsfs}
\usepackage[amsmath,thmmarks]{ntheorem}

\newcommand{\GeV}{\,\mathrm{GeV}}

\newcommand{\vev}[1]{\left\langle #1\right\rangle}

\newlength{\myem}
\newcounter{mysubequation}[equation]

\newcommand{\SISSA}{School of Physics and State Key Laboratory of Nuclear Physics and Technology, Peking University,  Beijing 100871, China \\
}

\newcommand{\preprintdate}{\hfill}
\newcommand{\preprintnumber}{%?????--??/2010/
}

\newcommand{\titletext}{Gauge Mediation of Supersymmetry Breaking\\
and the Froggatt-Nielsen Mechanism}
\newcommand{\authortext}{\large Zhong-hua Zhang and Da-Xin Zhang
\medskip\\\em\normalsize
\SISSA}
\newcommand{\abstracttext}{}

\title{
\normalsize
\begin{tabular}[t]{l}%\hepnumber\\
\preprintdate\end{tabular}
\hspace*{\fill}
\begin{tabular}[t]{l}\preprintnumber\end{tabular}
\vspace{3\baselineskip}\\\Large\bfseries\titletext\bigskip}
\author{\begin{minipage}[t]{0.90\textwidth}
\normalsize\centering\authortext
\end{minipage}}
\date{}

\begin{document}

\bigskip
\maketitle
\begin{abstract}\normalsize\noindent
\abstracttext In the model of gauge mediation of SUSY breaking in
the presence of tree-level mediation, the Froggatt-Nielsen mechanism
provides a different hierarchy of sparticle masses. We study the
spectra and show the results to be like those in an effective
supersymmetric model.
\end{abstract}\normalsize\vspace{\baselineskip}

% \clearpage

% \noindent
\section{}
Supersymmetry (SUSY) is an important candidate for the physics
beyond the Standard Model (SM). Among many models of SUSY breaking,
gauge mediation of SUSY breaking (GMSB) \cite{NPB189,PLB110,PRD48,PRD51,PRD53} provides an
important mechanism. In the usual GMSB models, SUSY is broken in a
hidden sector and the effects of SUSY breaking are transmitted to
the visible sector by gauge interactions of the SM gauge groups. It
was noticed in \cite{tree} that the effects of SUSY breaking can be
extended in the presence of extra gauge interactions. In \cite{tree}
this extra gauge group is a U(1)-subgroup of SO(10) in the grand
unification theory (GUT) and it is broken at around the unification
scale. Consequently, SUSY particles receive tree-level SUSY breaking
effects mediated by the extra U(1) gauge interactions if these
particles carry the U(1) quantum numbers. The supertrace formula constrains the total sfermion and the corresponding fermion total squared masses to be the same \cite{supertrace}. In this paper, the supertrace formula can be satisfied by compensating the contribution from MSSM particles by the heavy particles of the GUT scale \cite{tree}.

Here we will study the possibility that the extra gauged U(1) is the
flavor symmetry of the Froggatt-Nielsen mechanism (FNM) \cite{NPB147,Babu}
such that the mechanisms of SUSY breaking and of  generation of
fermion masses have a common origin. Different from the model of
\cite{tree}, the quantum numbers and consequently the spectra of the
SUSY particles are very different. There are superfields which do
not carry the U(1)$_{FN}$ quantum numbers, their scalar components
get masses at two-loop level.

In the literature, combining FNM and GMSB has been studied in
\cite{fng} where SUSY is broken at low energy ($\sim 100$TeV).
Unlike the original FNM where extra (super-)fields are at the GUT or
even higher scale, in \cite{fng} the extra superfields are
introduced at the low energy scale and they mix with those in the
minimal supersymmetric standard model (MSSM), especially in the
fermion sector. Consequently, unitarity of the $3\times 3$ quark
mixing matrix is violated, which makes the model rather complicated
\cite{fng2}. Different from \cite{fng}, in the present work the
extra superfields are introduced at around the GUT scale as in the
original FNM. At low energy the model is reduced to be the MSSM,
thus the possible problem of unitarity violation is absent.

We will firstly present the model. Then we will classify the
mechanism for generating the SUSY breaking parameters. Finally we
will present the spectra and summarize.

\section{}
The model extends the SM gauge group by adding a gauged U(1)$_{FN}$.
Instead of being a generator of GUT group, we take this U(1)$_{FN}$
to be irrelevant with unification as there exists no evidence of any
relevance. The GUT group is taken to be SU(5), although a larger
group is also possible.

At around the GUT scale we add in extra Higgs-like superfields and
matter-like superfields. The Higgs-like superfields are  those
complete representations of SU(5) as in the original model in
\cite{NPB147,Babu}, extended by a SU(5) singlet sector. Besides the possibly
existed colored Higgs whose effects are calculable, the newly
introduced Higgs-like superfields will not affect gauge coupling
unification. The singlet sector contains at least two effective
superfields: $N^\prime$ whose F-component has a nonzero vacuum
expectation value (VEV) which breaks SUSY,
\begin{equation}
\label{1} \vev{N'} = F\, \theta^2,
\end{equation}
and $N$ whose VEV breaks the U(1)$_{FN}$,
\begin{equation}
\label{Ns-vev} \vev{N} = M \sim M_{GUT}.
\end{equation}
The U(1)$_{FN}$ charges of both $N^\prime$ and $N$ are taken to be 5
for convenience. Note that we do not care if they are elementary or
composite.

The extra heavy matter-like superfields form complete
representations of the GUT SU(5) group so as not to spoil the gauge
coupling unification. For those which are relevant to SUSY breaking,
They are taken to be vector-like $F+\bar F=5+\bar 5,10+\overline{
10}+...$ . We take their U(1)$_{FN}$ charges to be $-3$ for $F$ and
$-2$ for $\bar F$. The superpotential of this heavy sector is
\begin{equation}
\label{w} W=hNF\bar F+h^\prime N^\prime F\bar F.
\end{equation}

The remaining chiral superfields are those in the MSSM. Their
U(1)$_{FN}$ charges\footnote{In some FN models the charges of u and e contain parameters p and s respectively, which can be different or the same. However, in our scheme, the $U(1)_{FN}$ anomalies must be canceled by the Green-Schwarz mechanism\cite{Green-Schwarz PLB149,Green-Schwarz NPB254,Green-Schwarz NPB255}, which requires p=s and that the FN charges in the same SU(5) multiplet should be the same.} are,
\begin{eqnarray}
\label{A} A_{Q_1,Q_2,Q_3}={4,2,0}, ~~A_{\bar u_1,\bar u_2,\bar
u_3}={4,2,0}, ~~A_{\bar d_1,\bar d_2,\bar d_3}={1+s,s,s},\\
~~A_{L_1,L_2,L_3}=1+s,s,s, ~~A_{\bar e_1,\bar e_2,\bar e_3}=4,2,0,
~~A_{H_u}=A_{H_d}=0,
\end{eqnarray}
as in the original FNM. The integer $s$ is allowed to take  0,1 or
2, corresponding to $\tan\beta$ taking a large, middle or small
value. Note that there exist no direct couplings between MSSM
superfields and $N$ or $N^\prime$ so that at low energy MSSM is
recovered.

\section{}
In the presence  of the nonzero VEV in (\ref{1}), the vector
superfield of the U(1)$_{FN}$ gauge group develops an effective
D-component as
\begin{equation}
\label{D} <D_{FN}>=2gA_{N^\prime}\bigg(\frac{F}{M_V}\bigg)^2,
\end{equation}
where $M_V$ is the mass of the gauge field of the U(1)$_{FN}$ group.
The other heavy superfields are the $F+\bar F$'s, they receive SUSY
breaking effect through (\ref{w}).

The effects of SUSY breaking are transmitted to the MSSM sector in
the following ways:

(1) The scalars of the MSSM chiral superfields which carry the
U(1)$_{FN}$ charges get masses at tree-level through couplings to
the U(1)$_{FN}$ vector superfield,
\begin{equation}%
\label{tree}
%\label{sfermions}
\tilde m^2_F = \frac{A_F}{2A_N}\, m^2,  \qquad m \equiv \frac{F}{M}.
\end{equation}

(2) The gauginos of the MSSM acquire masses in the same way as in
the usual GMSB model, where again the heavy $F+\bar F$ play the
roles of messengers,
\begin{eqnarray}
\label{gauginos1} M_a = \frac{\alpha_a}{4\pi} m \sum_{i,j}
\frac{h'_{ij}}{h_{ij}}n_a(i,j)g(x_{ij}),\quad x_{ij}
=\frac{h'_{ij}F}{h_{ij}M},\quad a=1,2,3,\\
g(x)=\frac{1}{x^2}[(1+x)\log(1+x)+(1-x)\log(1-x)].
\end{eqnarray}

(3) The scalars of the MSSM chiral superfields which do not carry
the U(1)$_{FN}$ charges get masses at 2-loop level using the same
mechanism as in the usual model of GMSB, where the heavy $F+\bar F$
play the roles of messengers,
\begin{equation}\label{2loop1}
\tilde m^2(Q)_{2loop} = 2\, \eta \, \sum_{a}c_a(Q) M^2_{a}, \qquad
\eta = \frac{\sum (h'_i/h_i)^2}{(\sum_i h'_i/h_i)^2} \geq
\frac{1}{3}.
\end{equation}
At the same time, those scalars in (\ref{tree}) receive corrections
(\ref{2loop1}), accordingly.

\section{}
Now we are ready to calculate the spectrum in the MSSM. We will
first input the value $F$, $M$, $\Lambda_{GUT}$. Then we will take eq.\eqref{2loop1}
as initial values at the GUT scale. Finally, we will use the
renormalization group equations to run the scale down to the
electroweak scale. The final step will be done with the help of {\tt
Suspect\,2.41}\cite{Djouadi}.

Numerically the unified gaugino mass is taken as $M_{1/2}=300Gev$.
For the typical scale of sparticle mass we take $m=3.2 TeV$. The
lightest neural Higgs mass is taken to satisfy the bound
$m_h>114GeV$. We take $s=0,1,2$ for $\tan\beta=30, 20, 9$,
respectively. We give these spectra  in Table \ref{spe30},
\ref{spe15}, \ref{spe9}, respectively.
% The spectrum for  $\tan\beta=30,~s=0$ is also plot in

\begin{table}[t]
\begin{center}
%\begin{center}
\begin{minipage}[]{0.45\linewidth}
\begin{tabular}{|c|c|c|c|c|}
\hline
Names &  & s=0 & s=1 &s=2\\
\hline\hline
Higgs: & $m_{h^0}$& 115.0 & 115.1 &115.2\\
&$m_{H^0}$&1357&1357&1357\\
&$m_A$&1357&1357&1357\\
&$m_{H^{\pm}}$&1359&1359&1359\\
\hline
Gluinos: & $M_{\tilde g}$ & 494.0 & 501.6&505.4\\
\hline
Neutralinos:&$m_{\chi_1^0}$& 125.4&125.4&125.4\\
&$m_{\chi_2^0}$& 231&231&231\\
&$m_{\chi_3^0}$& 840&840&840\\
&$m_{\chi_4^0}$& 843&843&843\\
\hline
Charginos:&$m_{\chi_1^\pm}$&231&231&231\\
&$m_{\chi_2^\pm}$&844&844&844\\
\hline
Squarks: &$m_{\tilde u_L}$&2159&2159&2159\\
&$m_{\tilde u_R}$&2127&2127&2127\\
&$m_{\tilde d_L}$&2160&2160&2160\\
&$m_{\tilde d_R}$&1183&1183&1183\\
&$m_{\tilde c_L}$&1605&1605&1605\\
&$m_{\tilde c_R}$&1562&1562&1562\\
&$m_{\tilde s_L}$&1607&1607&1607\\
&$m_{\tilde s_R}$&604&1183&1562\\
&$m_{\tilde t_1}$&625&627&630\\
&$m_{\tilde t_2}$&731&732&733\\
&$m_{\tilde b_1}$&596&706&707\\
&$m_{\tilde b_2}$&715&1185&1563\\
\hline
Sleptons:
&$m_{\tilde e_L}$&1127&1514&1821\\
&$m_{\tilde e_R}$&2051&2051&2051\\
&$m_{\tilde \mu_1}$&495&1127&1514\\
&$m_{\tilde \mu_2}$&1469&1469&1469\\
&$m_{\tilde \tau_1}$&322&329&330\\
&$m_{\tilde \tau_2}$&502&1127&1515\\
&$m_{\tilde \nu_e}$&1124&1512&1820\\
&$m_{\tilde \nu_\mu}$&489&1124&1512\\
&$m_{\tilde \nu_\tau}$&489&1124&1512\\
\hline
\end{tabular}
\caption{The spectra corresponding to $\tan\beta =30$,with s=0,1,2 respectively. All the masses are in GeV. \label{spe30}}
%\vspace{-0.45cm}
\end{minipage}
\end{center}
\end{table}
%
%\end{center}

\begin{table}[t]
\begin{center}
\begin{minipage}[]{0.45\linewidth}
\begin{tabular}{|c|c|c|c|c|}
\hline
Names &  & s=0 & s=1 &s=2\\
\hline\hline
Higgs: & $m_{h^0}$& 114.5 & 114.6 &114.6\\
&$m_{H^0}$&1356.5&1356.5&1356.5\\
&$m_A$&1356.5&1356.5&1356.5\\
&$m_{H^{\pm}}$&1359&1359&1359\\
\hline
Gluinos: & $M_{\tilde g}$ & 494.0 & 502.0&505.4\\
\hline
Neutralinos:&$m_{\chi_1^0}$& 125&125&125\\
&$m_{\chi_2^0}$& 230.4&230.4&230.4\\
&$m_{\chi_3^0}$& 839.5&839.5&839.5\\
&$m_{\chi_4^0}$& 843&843&843\\
\hline
Charginos:&$m_{\chi_1^\pm}$&230&230&230\\
&$m_{\chi_2^\pm}$&844&844&844\\
\hline
Squarks: &$m_{\tilde u_L}$&2159&2159&2159\\
&$m_{\tilde u_R}$&2127&2127&2127\\
&$m_{\tilde d_L}$&2160&2160&2160\\
&$m_{\tilde d_R}$&1183&1562&1866\\
&$m_{\tilde c_L}$&1605&1605&1605\\
&$m_{\tilde c_R}$&1562&1562&1562\\
&$m_{\tilde s_L}$&1607&1607&1607\\
&$m_{\tilde s_R}$&604&1183&1562\\
&$m_{\tilde t_1}$&625&627&630\\
&$m_{\tilde t_2}$&732&732&733\\
&$m_{\tilde b_1}$&599&708&708\\
&$m_{\tilde b_2}$&713&1183&1562\\
\hline
Sleptons:
&$m_{\tilde e_L}$&1127&1514&1821\\
&$m_{\tilde e_R}$&2051&2051&2051\\
&$m_{\tilde \mu_1}$&495&1127&1514\\
&$m_{\tilde \mu_2}$&1469&1469&1469\\
&$m_{\tilde \tau_1}$&326&331&331\\
&$m_{\tilde \tau_2}$&499&1127&1514\\
&$m_{\tilde \nu_e}$&1124&1512&1820\\
&$m_{\tilde \nu_\mu}$&489&1124&1512\\
&$m_{\tilde \nu_\tau}$&489&1124&1512\\
\hline
\end{tabular}
\caption{The spectra corresponding to $\tan\beta =15$, with s=0,1,2 respectively. All the masses are in GeV.  \label{spe15}}
%\vspace{-0.45cm}
\end{minipage}
\end{center}
\end{table}

\begin{table}[t]
\begin{center}
\begin{minipage}[]{0.45\linewidth}
\begin{tabular}{|c|c|c|c|c|}
\hline
Names &  & s=0 & s=1 &s=2\\
\hline\hline
Higgs: & $m_{h^0}$& 113.4 & 113.5 &113.5\\
&$m_{H^0}$&1356.7&1356.7&1356.7\\
&$m_A$&1356.7&1356.7&1356.7\\
&$m_{H^{\pm}}$&1359&1359&1359\\
\hline
Gluinos: & $M_{\tilde g}$ & 493.9 & 501.5&505.4\\
\hline
Neutralinos:&$m_{\chi_1^0}$& 125&125&125\\
&$m_{\chi_2^0}$& 229.7&229.7&229.7\\
&$m_{\chi_3^0}$& 839&839&839\\
&$m_{\chi_4^0}$& 844&844&844\\
\hline
Charginos:&$m_{\chi_1^\pm}$&229.7&229.7&229.7\\
&$m_{\chi_2^\pm}$&844.6&844.6&844.6\\
\hline
Squarks: &$m_{\tilde u_L}$&2159&2159&2159\\
&$m_{\tilde u_R}$&2127&2127&2127\\
&$m_{\tilde d_L}$&2160&2160&2160\\
&$m_{\tilde d_R}$&1183&1562&1866\\
&$m_{\tilde c_L}$&1605&1605&1605\\
&$m_{\tilde c_R}$&1562&1562&1562\\
&$m_{\tilde s_L}$&1607&1607&1607\\
&$m_{\tilde s_R}$&604&1183&1562\\
&$m_{\tilde t_1}$&623&625&628\\
&$m_{\tilde t_2}$&733&733&733\\
&$m_{\tilde b_1}$&602&709&709\\
&$m_{\tilde b_2}$&710&1183&1562\\
\hline
Sleptons:
&$m_{\tilde e_L}$&1127&1514&1821\\
&$m_{\tilde e_R}$&2051&2051&2051\\
&$m_{\tilde \mu_1}$&489&1127&1514\\
&$m_{\tilde \mu_2}$&1469&1469&1469\\
&$m_{\tilde \tau_1}$&330&331&331\\
&$m_{\tilde \tau_2}$&496&1127&1514\\
&$m_{\tilde \nu_e}$&1124&1512&1820\\
&$m_{\tilde \nu_\mu}$&489&1124&1512\\
&$m_{\tilde \nu_\tau}$&489&1124&1512\\
\hline
\end{tabular}
\caption{The spectra corresponding to $\tan\beta =9$, with s=0,1,2 respectively. All the masses are in GeV. \label{spe9}}
\vspace{-0.45cm}
\end{minipage}
\end{center}
\end{table}

\begin{figure}[h]
\centering
\includegraphics[width=9.5cm,height=12cm]{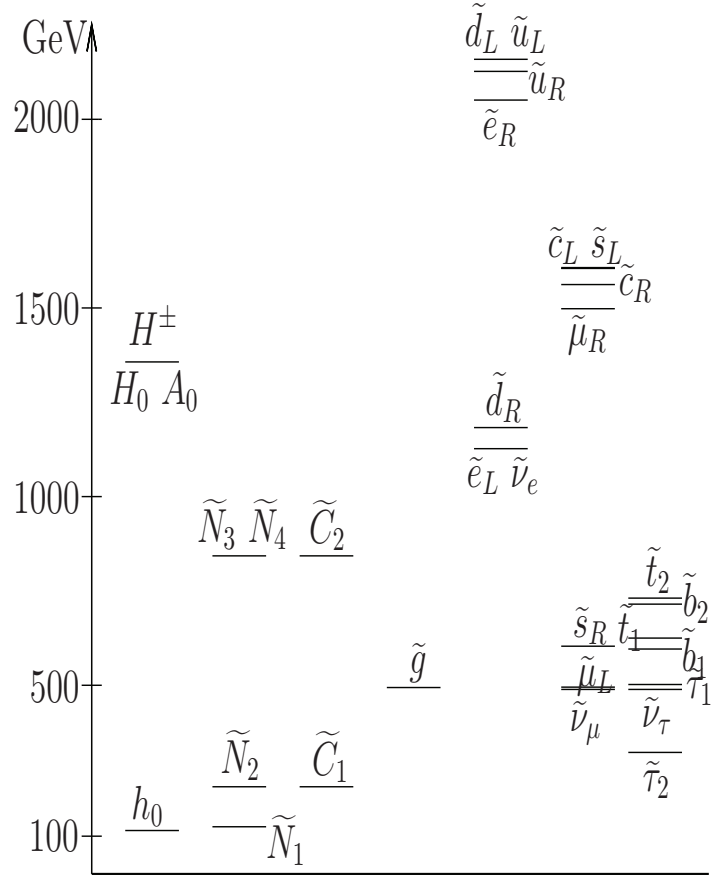}
\caption{An example of spectrum, corresponding to $m=3.2 TeV,
M_{1/2} = 300 Gev, \tan \beta = 30, s=0 ,\text{sign}(\mu) = +, A=0,
\eta = 1$. All the masses are in $\GeV$.} \label{spectrum}
\end{figure}

We find that the NLSP is always a Bino-like neutralino whose mass is
around $125$GeV. The mass of the lighter chargino is around
$230$GeV, which is very different from the result of nearly
degeneration of lightest neutralino and the lighter chargino in
\cite{tree}. These results imply that they can be easily produced
and identified in the LHC experiments. We find, also different from
the results of \cite{tree} where all the sfermions are heavy, that
the sfermions for the first two generations are usually heavy,
corresponding to the effective SUSY scenario\cite{effective} of suppressing
FCNC. The sfermions of the third generation are generally lighter
than 1TeV, making them accessible at the LHC experiments. To be
intuitive, we also plot an example in Fig. \ref{spectrum}.

\section{}
In summery, we have studied the spectra in the model of GMSB with
FNM. We find the results are very different from those in
\cite{tree}. No tree-level FCNC is introduced due to a high energy
scale of SUSY breaking, which is different from the case in
\cite{fng,fng2}.

This work was supported in part by the National Natural Science
Foundation of China (NSFC) under the grant No. 10435040.

\end{document}